\documentclass[a4paper]{llncs}

\usepackage{inputenc}
\usepackage{graphicx}
\usepackage{amssymb}
\usepackage{amsmath}
\usepackage{makecell}
\usepackage[figuresleft]{rotating}

\usepackage{color}

\usepackage[british]{babel}
\usepackage{todonotes}

\usepackage{caption}
\usepackage{multirow}
\usepackage{url}

\usepackage[backend=biber,style=ieee, maxnames=3]{biblatex}
\usepackage{csquotes}

\usepackage{acronym}

\begin{document}

\mainmatter  

\title{Sparse annotation strategies for segmentation of short axis cardiac MRI}

\titlerunning{}


\author{Josh Stein \inst{1, 2}, Maxime Di Folco \inst{1} \and Julia A. Schnabel \inst{1,2,3}}
\authorrunning{J. Stein et al.}

%

\institute{ Institute of Machine Learning in Biomedical Imaging, Helmholtz Munich, Neuherberg, Germany
            \and 
            Technical University of Munich, Munich, Germany
            \and 
            King’s College London, London, UK
            \\ \textbf{Preliminary work. Under review.}}


\maketitle

\begin{abstract}

Short axis cardiac MRI segmentation is a well-researched topic, with excellent results achieved by state-of-the-art models in a supervised setting. However, annotating MRI volumes is time-consuming and expensive. Many different approaches (e.g. transfer learning, data augmentation, few-shot learning, etc.) have emerged in an effort to use fewer annotated data and still achieve similar performance as a fully supervised model. Nevertheless, to the best of our knowledge, none of these works focus on \emph{which} slices of MRI volumes are most important to annotate for yielding the best segmentation results. In this paper, we investigate the effects of training with sparse volumes, i.e. reducing the number of cases annotated, and sparse annotations, i.e. reducing the number of slices annotated per case. We evaluate the segmentation performance using the state-of-the-art nnU-Net model on two public datasets to identify which slices are the most important to annotate. We have shown that training on a significantly reduced dataset (48 annotated volumes) can give a Dice score greater than 0.85 and results comparable to using the full dataset (160 and 240 volumes for each dataset respectively). In general, training on more slice annotations provides more valuable information compared to training on more volumes. Further, annotating slices from the middle of volumes yields the most beneficial results in terms of segmentation performance, and the apical region the worst. When evaluating the trade-off between annotating volumes against slices, annotating as many slices as possible instead of annotating more volumes is a better strategy.

\end{abstract}

\section{Introduction}
\label{sec:Intro}




Cardiac image segmentation constitutes a fundamental initial phase in various applications. Segmentation is often the first step in evaluating cardiac functionality in order to diagnose disease. By partitioning the image into distinct semantically meaningful regions, typically aligned with anatomical structures, it facilitates the extraction of quantitative measures critical for further analysis and interpretation \cite{chen2020deep}. Deep learning methods have become the state-of-the-art approach for this task, but they require the collection and annotation of data, which are time-consuming and laborious processes. 

Much effort has been spent improving methods that require fewer ground truth annotations. Some popular approaches include data augmentation \cite{zhao2019data}, transfer learning \cite{bai2018automated,khened2019fully,chen2019med3d}, semi-supervised learning \cite{qin2018joint,can2018learning}, and self-supervised learning \cite{bai2019self,zeng2021positional} - many others exist \cite{peng:2021}.

In general, the data used for these methods are \emph{sparse} or \emph{limited}. The exact definition varies from context to context and usually falls into one of three broad categories \cite{peng:2021}. First, the annotation of data volumes could be sparse, where only particular patients may be annotated. Second, the annotation of slices within volumes could be sparse. Instead of having a fully annotated 3D volume, only particular slices within the volume may be annotated. Third, there could be sparsity in the slice annotations themselves. Instead of having a pixel-accurate ground truth annotation, there may be bounding boxes, scribbles, or particular labelled points. These categories are not mutually exclusive - for example, there may be sparse slices that are sparsely annotated.

In cardiac imaging, segmentation of the short-axis view on MRI data has been well studied, thanks to technical challenges \cite{bernard2018deep,campello2021multi}. Nowadays, we consider fully supervised short-axis cMRI segmentation a well-researched task, with state-of-the-art approaches surpassing human performance.


In this paper, we purposely choose not to use any approaches for sparse data to answer the following questions, which are still unclear in the literature:

\begin{enumerate}
    \item How many sparse data does are needed for achieving reasonable results with state-of-the-art network nnU-Net?
    \item Which cardiac regions (basal, mid or apical) contribute the most to segmentation performance?
    \item Is there a particular annotation strategy which one should prefer between annotating volumes or annotating slices?
\end{enumerate}

In this work, we investigate the effects on segmentation performance on two public datasets \cite{bernard2018deep,campello2021multi}, when removing volumes (reducing the number of annotated case), removing slice annotations (both randomly and from particular cardiac regions) and the balance between these two.

\section{Related work}

Recent works on cardiac imaging segmentation have focused on using sparse annotations while still achieving results that compare to using a fully annotated dataset. Bitarafan et al. \cite{Bitarafan_Nikdan_Baghshah_2021} use a single annotated 2D slice with registration and self-training to propagate and train on label propagations. They achieve approximately a 10\% reduction in Dice score using a single annotation compared to using a fully annotated volume. Bai et al. \cite{Bai_Suzuki_Qin_Tarroni_Oktay_Matthews_Rueckert_2018} also use label propagation in combination with a recurrent neural network (RNN) to incorporate both spatial and temporal information. Using two annotated frames, they are able to out-compete a baseline U-Net model (trained on all available annotated frames). Contrastive learning strategies have also been explored. Zeng et al. \cite{zeng2021positional} use a contrastive loss between slice position in a self-supervised pre-training stage that achieves a similar Dice score to a fully annotated dataset using only 15 annotated volume (the achieved Dice score is 3\% lower compared to using the fully annotated set). You et al. \cite{you2022mine} present a contrastive semi-supervised 2D medical segmentation framework for very limited annotations and accomplish a Dice score of 0.82 using only 1\% of labels and similar performance to fully annotated set using only 10\% of the labels. 

\section{Methods}
\label{sec:Methods}

\subsection{Segmentation network}

We use nnU-Net \cite{isensee2021nnu}, the current state-of-the-art model for cardiac segmentation (achieving first place in ACDC and M\&Ms challenges). We do not modify the standard nnU-Net processing pipeline, except to change the data sampling strategy. We evaluate on mean foreground Dice score, Hausdorff Distance (HD) and Mean Absolute Distance (MAD).

We evaluate both 3D and 2D nnU-Net models. For both datasets, we evaluate only the 3D high resolution model (i.e. not the low resolution or cascaded model). The reader is referred to the original nnU-Net paper \cite{isensee2021nnu} for clarification on the differences between these models and their processing pipelines. The 2D models are trained by sampling a slice from a given volume.

\subsection{Definition of sparsity}

For each dataset, we investigate the effects of removing volumes (i.e. enforcing sparsity of volumes), zeroing out slices (i.e. enforcing sparsity of slices) and the balance between these two.

\paragraph{\textbf{Sparse volumes:}} To investigate the sparsity of volumes, we randomly sample a percentage of the total patient cardiac volumes, which are then used for training. By iteratively training on smaller samples of the dataset we want to determine the number of needed annotated volumes to achieve comparable results to training on the full dataset. 

\paragraph{\textbf{Sparse slices}:} We investigate slice sparsity by sampling and training on several slices from within a volume. All non-sampled slices are zeroed, which allows us to maintain volume shape. This prevents the need to modify network parameters, which would otherwise need to continuously adapt based on the input volume. Slices can either be sampled randomly (from within the entire volume) or explicitly sampled from the apical, mid-ventricular, and/or basal regions. We assume that each volume is split into equal thirds, where the first third corresponds to apical slices, the second mid-ventricular slices and the third basal slices. Sampling from various permutations of these regions allows us to investigate which (if any) regions are most important for segmentation performance.

\paragraph{\textbf{Annotation strategy}: } Finally, we investigate the balance and relative importance of sparse volumes vs sparse slices by sampling a percentage of the volumes and then randomly sampling different percentages of slices from within the sampled volumes.

\section{Experiments and results}
\label{sec:Results}

\subsection{Dataset} 

We use the Automatic Cardiac Diagnosis Challenge (ACDC) \cite{bernard2018deep} and the Multi-Centre, Multi-Vendor and Multi-Disease Cardiac Segmentation Challenge (M\&Ms) \cite{campello2021multi}. These datasets are popular in the literature, and have been used to investigate a variety of supervised and unsupervised segmentation methods. They are both inherently sparse (across volumes) as they provide annotations at only end-diastolic and end-systolic phases.

ACDC has a set of 100 training cases, each of which has two fully annotated volumes (one at end-diastole, one at end-systole). We train on all 200 volumes using 160 volumes for training and the remaining 40 volumes for validation. The test set is composed of 100 cases, each of which is again fully annotated at end-diastole and end-systole. After nnU-Net preprocessing transformations, all volume have 20 slices. The total number of available training slices is therefore $20 \times 160 = 3200$.

Similarly, M$\&$Ms has a set of 150 training cases, each of which has end-diastole and end-systole volumes annotated. We split the training data into 240 training cases and 60 validation cases. The test set is composed of 136 cases (again, each case has end-diastole and end-systole annotated).
 After nnU-Net preprocessing transformations, there are 14 slices per volume - the total number of available training slices is therefore $14 \times 240 = 3360$. The M$\&$Ms dataset contains images from different vendors and centers. We keep the same split of dataset described in \cite{campello2021multi}. Please refer to the corresponding paper for details on the acquisition protocol.

\subsection{Sparse volumes}

The results of using a reduced dataset for both ACDC and M\&Ms are shown in Table \ref{tab:reduced_data}. First, we observe that the networks trained on ACDC generally outperform those trained on M\&Ms. We believe this is due to the different MRI domains present in the M\&Ms dataset. Second, we note that using more than 48 volumes (approximately 30\% of the ACDC dataset, and 20\% of the M\&Ms dataset), regardless of dataset or model dimensionality, yields a Dice score greater than 0.85. However, using fewer volumes leads to increases in corresponding HD and MAD scores, and a decrease in Dice scores. Further, using fewer than 48 volumes leads to a worse performance in networks trained on ACDC data compared to those trained on M\&Ms. However, when the number of volumes is severely restricted (e.g. using 8 volumes) we see a similarly poor performance for both datasets. Finally, we note that the difference in performance between 2D and 3D networks is more pronounced for networks trained on ACDC than those trained on M\&Ms. This is especially true when considering the difference in surface distance metrics between 2D and 3D networks.
We conclude that having a variety of domains within the M\&Ms dataset makes it more difficult to achieve higher Dice scores, while simultaneously allowing for better generalisation when removing annotated volumes.

\begin{table}[ht]
\centering
\resizebox{1\textwidth}{!}{\begin{tabular}{|c|c|c|c|c|c|c|c|c|c|c|c|}
\hline
 &  &  & \multicolumn{8}{c|}{Number of training volumes}  \\
 \hline
Network & Dataset & Evaluation metric & 8 & 24 & 32 & 48 & 80 & 160 & 192 & 240  \\
\hline
\multirow{3}{*}{2D nnU-Net} & \multirow{3}{*}{ACDC} & Dice & 0.62 & 0.71 & 0.74 & 0.85 & 0.89 & \textbf{0.91} & - & - \\
 &  & HD (mm) & 36.03 & 22.15 & 18.56 & 14.61 & 7.20 & \textbf{5.06} & - & -  \\
 &  & MAD (mm) & 10.16 & 5.67 & 8.95 & 2.81 & 1.65 & \textbf{1.16} &  - & - \\
 \hline
\multirow{3}{*}{3D nnU-Net} & \multirow{3}{*}{ACDC} & Dice & 0.57 & 0.66 & 0.71 & 0.85 & 0.85 & \textbf{0.91} & - & -   \\
 &  & HD (mm) & 54.93 & 43.63 & 29.34 & 8.81 & 7.75 & \textbf{4.4} & - & -  \\
 &  & MAD (mm) & 18.03 & 13.61 & 8.95 & 2.17 & 1.94 & \textbf{1.16} & - & - \\
\hline
\multirow{3}{*}{2D nnU-Net} & \multirow{3}{*}{M\&Ms} & Dice & 0.6 & 0.82 & 0.83 & 0.85 & 0.85 & 0.86 & 0.86 & \textbf{0.87}  \\
 &  & HD (mm) & 32.43 & 9.3 & 9.42 & 8.81 & 7.75 & 6.84 & 6.87 & \textbf{6.54}  \\
 &  & MAD (mm) & 8.9 & 2.39 & 2.38 & 2.17 & 1.94 & 1.74 & 1.74 & \textbf{1.74} \\ 
 \hline
\multirow{3}{*}{3D nnU-Net} & \multirow{3}{*}{M\&Ms} & Dice & 0.54 & 0.82 & 0.82 & 0.84 & 0.85 & 0.86 & 0.86 & \textbf{0.87}  \\
 &  & HD (mm) & 37.41 & 9.1 & 8.86 & 6.98 & 6.44 & 5.89 & \textbf{5.8} & 6.02  \\
 &  & MAD (mm) & 11.26 & 2.25 & 2.32 & 1.79 & 1.65 & 1.56 & \textbf{1.53} & 1.6 \\ 
 \hline
\end{tabular}}

\caption{Effect of training on sparse annotated volumes. Note that the ACDC dataset only has 160 volumes.}
\label{tab:reduced_data}
\end{table}

\subsection{Sparse annotations}

The results of training 3D nnU-Net only on particular cardiac regions are shown in Table \ref{tab:reduced_slice_regions}. As expected, the best performance is achieved by using all three cardiac regions (i.e. the most slices). Using only two regions, all combinations achieve a Dice score greater than 0.8, except for the network trained on the apical and basal combination trained on M\&Ms data, which achieves 0.79. We also observe that training on combinations using mid-ventricular slices (i.e. apical and mid, or basal and mid combinations) yield the best-performing networks. The worst performance is achieved by networks trained on only a single cardiac region (the network trained on M\&Ms data trained solely on the apical region is a very poor-performing network). The best-performing single region networks are those trained on the basal region for ACDC data and the middle region for M\&Ms data.

\begin{table}[ht]
\centering
\resizebox{1\textwidth}{!}
{\begin{tabular}{|c|c|c|c|c|c|c|c|c|c|}
\hline
 &  & \multicolumn{7}{c|}{Cardiac regions trained on}   \\
 \hline
 Dataset & Metric & A + M + B & A + M & M + B & A + B & A & M & B  \\
\hline
\multirow{3}{*}{ACDC} & Dice & \textbf{0.91} & 0.88 & 0.89 & 0.81 & 0.54 & 0.5 & 0.69 \\
 & HD (mm) & \textbf{4.17} & 6.66 & 5.89 & 52.47 & 21.04 & 149.41 & 26.37  \\
 & MAD (mm) & \textbf{1.08} & 1.96 & 1.76 & 18.45 & 6.35 & 64.59 & 8.16  \\
 \hline
\multirow{3}{*}{M\&Ms} & Dice & \textbf{0.87} & 0.84 & 0.86 & 0.79 & 0.08 & 0.69 & 0.38 \\
 & HD (mm) & \textbf{5.52} & 11.48 & 7.0 & 9.86 & 46.5 & 59.53 & 51.21 \\
 & MAD (mm) & \textbf{1.47} & 4.17 & 2.17 & 2.63 & 21.75 & 19.82 & 17.22 \\
 \hline
\end{tabular}}
\caption{Effect of training on sparse annotated slices from different cardiac regions (A=apical slices, M=middle slices, B=basal slices).}
\label{tab:reduced_slice_regions}
\end{table}

We then train networks on randomly sample slicesd from all three cardiac regions, the results of which are show in Table \ref{tab:random_reduced_slices}. The aforementioned cardiac regions correspond to using one-third of all slices per region. For ACDC, a single cardiac region corresponds to sampling approximately 6 slices, and two cardiac regions to approximately 13 slices, and 5 slices and 10 slices for M\&Ms, respectively. 

Randomly sampling a third of the available slices yields better results than sampling from any single cardiac region (although there is still a similar drop in performance when using a limited number of slices). Randomly sampling two thirds of available slices yields similar results as sampling from either the apical and middle slices or the middle and basal slices, which in turn is similar to using the full set of available slices. We also observe a Dice score greater than 0.8 when only using approximately 40\% of slices (i.e. 8 slices for ACDC, 6 slices for M\&Ms) with a slight decrease for the surface distance metrics. A number of 10 slices annotated is sufficient to achieve a Dice superior to 0.85 for both datasets. This corresponds to half the slices annotated for ACDC and around 70\% for M\&Ms.

\begin{table}[ht]
\setlength\extrarowheight{6pt}
\centering

\resizebox{1\textwidth}{!}{\begin{tabular}{|c|c|c|c|c|c|c|c|c|c|c|}
\hline
 &  & \multicolumn{9}{c|}{Number of slices used for training} \\
\hline
 Dataset & Metric & 1 & 2 & 4 & 6 & 8 & 10 & 14 & 16 & 20 \\
\hline
 \multirow{3}{*}{ACDC} & Dice & 0.01 $\pm 0.01$ & 0.28 $\pm 0.08$ & 0.62 $\pm 0.14$ & 0.77 $\pm 0.07$ & 0.81 $\pm 0.07$ & 0.87 $\pm 0.05$ & 0.9 $\pm 0.02$ & 0.9 $\pm 0.02$ & \textbf{0.91 $\pm 0.02$} \\
   & HD (mm) & 82.18 $\pm 18.83$ & 24.79 $\pm 3.46$ & 19.34 $\pm 7.77$ & 10.83 $\pm 1.27$ & 8.58 $\pm 0.9$ & 6.88 $\pm 2.3$ & 5 $\pm 1.75$ & 4.78 $\pm 1.0$ & \textbf{4.63 $\pm 1.6$} \\
   & MAD (mm) & 33.09 $\pm 4.65$ & 8.34 $\pm 1.21$ & 6.18 $\pm 2.82$ & 2.95 $\pm 0.15$ & 2.39 $\pm 0.27$ & 1.71 $\pm 0.43$ & 1.3 $\pm 0.3$ & 1.22 $\pm 0.22$ & \textbf{1.2 $\pm 0.33$} \\
 \hline
 \multirow{3}{*}{M\&Ms} & Dice & 0.01 $\pm 0.01$ & 0.28 $\pm 0.1$ & 0.68 $\pm 0.11$ & 0.79 $\pm 0.08$ & 0.83 $\pm 0.06$ & 0.85 $\pm 0.04$ & \textbf{0.86 $\pm 0.03$} & - & - \\
  & HD (mm) & 76.06 $\pm 16.43$ & 28.2 $\pm 6.05$ & 14.1 $\pm 2.57$ & 11.11 $\pm 3.33$ & 7.58 $\pm 1.35$ & 6.33 $\pm 0.76$ & \textbf{5.64 $\pm 1.05$} & - & - \\
  & MAD (mm) & 35.3 $\pm 3.47$ & 10.31 $\pm 2.85$ & 4.27 $\pm 0.68$ & 3.82 $\pm 1.87$ & 2.4 $\pm 0.76$ & 1.67 $\pm 0.14$ & \textbf{1.49 $\pm 0.2$} & - & - \\
 \hline
\end{tabular}}
\caption{Influence of training with randomly sampled and sparsely annotated slices from all three cardiac regions using 3D nnU-Net. Note that there are only 14 slices per volume for the M\&Ms dataset.}
\label{tab:random_reduced_slices}
\end{table}

\subsection{Sparse dataset vs sparse annotations}

In this section, we investigate annotation strategies using different balances of reduced volume annotations and reduced slice annotations while keeping a fixed number of total slices annotated. Table \ref{tab:proportionality_1400} shows the results for approximately 1400 slices annotated across both datasets. For both networks, when keeping the total number of slices the same, better results are achieved when using more slices per volume. 

This is further observed in Tables \ref{tab:proportions_acdc} and \ref{tab:proportions_mnms} where we compare approximately 700 annotated slices per dataset. Again, we note how best performance is achieved with more slices, even if a smaller number of volumes is annotated. We observe that in general, the networks trained on ACDC perform better than those trained on M\&Ms when using the same number of slices and volumes. Note that since the overall number of slices is quite similar for both datasets (3200 for ACDC, 3360 for M\&Ms) the relative proportion for a given slice/volume trial is also similar. Finally, we note that ACDC seems to be more affected by using fewer volumes - that is, the drop in performance when halving the number of volumes (and keeping the number of slices fixed) is slightly larger compared to M\&Ms. Despite this, we see the same overall pattern that using fewer slices leads to worse performance.

\begin{table}[ht]
\centering
\resizebox{1\textwidth}{!}{\begin{tabular}{|c|c|c|c|c|c|c|}
\hline
 & & \multicolumn{5}{c|}{Proportionality constant C $\times$ S = $\sim$1400} \\
 \hline
Dataset & Metric & 65 C, 20 S & 100 C, 14 S & 120 C, 12 S & 160 C, 9 S & 240 C, 6 S \\
\hline
\multirow{3}{*}{ACDC} & Dice & 0.87 $\pm 0.02$ & \textbf{0.89 $\pm 0.02$} & 0.88 $\pm 0.03$ & 0.82 $\pm 0.09$ &  \\
 & HD (mm) & 8 $\pm 1.47$ & 8.52 $\pm 0.72$ & \textbf{6.7 $\pm 0.5$} & 8.27 $\pm 1.45$&  \\
 & MAD (mm) & 1.98 $\pm 0.3$ & 2.08 $\pm 0.18$ & \textbf{1.6 $\pm 0.08$} & 2.01 $\pm 0.33$ &  \\
 \hline
\multirow{3}{*}{M\&Ms} & Dice &  & \textbf{0.86 $\pm 0.03$} & 0.84 $\pm 0.02$ & 0.83 $\pm 0.06$ & 0.77 $\pm 0.08$ \\
 & HD (mm) &  &  \textbf{6.1 $\pm 0.95$} & 6.72 $\pm 1.25$ & 6.93 $\pm 0.59$ & 9.28 $\pm 0.78$ \\
 & MAD (mm) &  &  \textbf{1.58 $\pm 0.16$} & 1.86 $\pm 0.15$ & 1.91 $\pm 0.22$ & 2.75 $\pm 0.23$ \\
 \hline
\end{tabular}}

\caption{Segmentation performance when changing the number of training cases vs the number of slices while keeping the total number of slices approximately 1400. C=cases, S=slices. Trained with a 3D nnU-Net network. Note that the M\&Ms dataset has a total of 14 slices per volume, and that the ACDC dataset has a total of 160 cases.}
\label{tab:proportionality_1400}
\end{table}

\begin{table}[t]
\centering
\resizebox{0.9\textwidth}{!}{\begin{tabular}{|c|c|c|c|c|c|c|c|c|c|c|c|c|}
 \hline
\multicolumn{1}{|r|}{Num slices} & \multicolumn{2}{c|}{20} & \multicolumn{2}{c|}{17} & \multicolumn{2}{c|}{14} & \multicolumn{2}{c|}{12} & \multicolumn{2}{c|}{10} & \multicolumn{2}{c|}{9} \\ \hline
Num cases & 65 & 32 & 80 & 40 & 100 & 50 & 120 & 60 & 144  & 77  & 160  & 80  \\ \hline
Dice & 0.87 & 0.7 & \textbf{0.89} & 0.84 & 0.87 & 0.87 & 0.88 & 0.87 & 0.87 & 0.85 & 0.83 & 0.78 \\
HD (mm) & 8 & 32.4 & \textbf{6.15} & 11.35 & 8.52 & 8.83 & 6.7 & 9.21 & 7.67 & 8.5 & 8.27 & 9.61 \\
MAD (mm) & 1.98 & 9.79 & \textbf{1.51} & 3.59 & 2.08 & 2.35 & 1.6 & 2.18 & 1.81 & 2.16 & 2.01 & 2.41 \\ \hline
\end{tabular}}
\caption{Influence of keeping slices constant while reducing number of training volumes. Trained on ACDC with a 3D nnU-Net network. Standard deviation scores are removed for brevity.}
\label{tab:proportions_acdc}
\end{table}

\begin{table}[t]
\centering
\resizebox{0.9\textwidth}{!}{\begin{tabular}{|c|c|c|c|c|c|c|c|c|c|c|c|c|}
\hline
Num slices & \multicolumn{2}{c|}{14} & \multicolumn{2}{c|}{12} & \multicolumn{2}{c|}{10} & \multicolumn{2}{c|}{9} & \multicolumn{2}{c|}{8} & \multicolumn{2}{c|}{6} \\
\hline
Num cases & 100 & 50 & 120 & 60 & 144 & 77 & 160 & 80 & 192 & 96 & 240 & 120 \\
\hline
Dice & \textbf{0.86} & 0.84 & 0.84 & 0.85 & 0.84 & 0.83 & 0.84 & 0.82 & 0.82 & 0.81 & 0.77 & 0.75 \\
HD (mm) & \textbf{6.1} & 7.14 & 6.72 & 6.3 & 7.07 & 6.92 & 6.93 & 7.79 & 8.17 & 7.77 & 9.28 & 12.11 \\
MAD (mm) & \textbf{1.58} & 1.79 & 1.86 & 1.67 & 1.93 & 1.83 & 1.91 & 2.22 & 2.34 & 2.15 & 2.75 & 3.76 \\
\hline
\end{tabular}}
\caption{Influence of keeping slices constant while reducing number of training volumes. Trained on M\&Ms with a 3D nnU-Net network. Standard deviation scores are removed for brevity.}
\label{tab:proportions_mnms}
\end{table}

\section{Discussion and Conclusion}

In this paper, we investigate the influence of different sparse annotation strategies for the segmentation of short-axis view in cardiac MRI. We use the state-of-the-art nnU-Net on two public datasets and evaluate using standard segmentation metrics. We show that good segmentation results can be achieved even when using a severely restricted dataset. We note that using more than 48 volumes is sufficient to achieve a Dice score greater than 0.85, and using more than 80 volumes is comparable to using the full dataset (160 volumes in ACDC, 240 volumes in M\&Ms). This corresponds to using half of all available volumes for ACDC (a total of 1600 slices), and one-third of available volumes for M\&Ms (a total of 1120 slices). 

Further, experiments with sparse annotations demonstrate that using more than two-thirds of available slices yields results comparable to using the full set of available annotations. We observe that randomly sampling these slices from throughout all cardiac regions results in better performance than sub-sampling from particular regions. If two regions are sub-sampled, the middle region contributes the most to segmentation performance and allows better generalisation. The conclusion differs between the dataset when we sub-sampled one region. Nevertheless, as we expect, the apical region generalises (and performs) the worst due to differences in ventricular sizes compared to the middle and basal regions.

Finally,  we demonstrate the importance of using more slices relative to volumes. When we use the full set of available training volumes with a limited number of slices, we achieve only poor results. However, even when the number of volumes is reduced, good performance can still be achieved if there is a large number of slices to learn from. For both datasets, annotating upwards of 60\% of the slices provides the best results. Therefore, we recommend annotating as many slices as possible in each volume instead of annotating more volumes with fewer slices. 

Future work will build on this baseline using state-of-the-art nnU-Net and compare different approaches for sparse annotations, such as transfer learning or semi-supervised learning, to evaluate the most appropriate strategy. Further, the study will be extended to include more datasets with different cardiac MRI views or modalities (e.g. ultrasound, CT).







\newpage
\footnotesize
\printbibliography
\end{document}